\documentclass[11pt,a4paper]{article}

\usepackage[truedimen,margin=34mm]{geometry} 

\usepackage{mathrsfs}
\usepackage{amssymb}
\usepackage{amsmath}
\usepackage{ascmac}
\usepackage{amsthm}
\usepackage[pdftex]{graphicx}
\usepackage[pdftex]{color}
\usepackage{booktabs}
\usepackage{setspace}
\usepackage{natbib}
\usepackage{color}

\usepackage{titlesec}
\titleformat*{\section}{\large\bfseries}
\titleformat*{\subsection}{\it}

\def\de{{\delta}}
\def\ep{{\varepsilon}}
\def\la{{\lambda}}

\def\ut{\widetilde{u}}

\def\ut{\widetilde{u}}

\def\st{\widetilde{s}}

\def\Re{\mathbb{R}}

\def\diag{\text{diag}}

\title{{\bf Estimation and Inference for Area-wise Spatial Income Distributions from Grouped Data}}

\date{}
\author{
}

\begin{document}

\maketitle
\doublespacing

\vspace{-1.5cm}
\begin{center}
Shonosuke Sugasawa$^1$, Genya Kobayashi$^2$ and Yuki Kawakubo$^2$
\end{center}

\noindent
$^1$Center for Spatial Information Science, The University of Tokyo\\
$^2$Graduate School of Social Sciences, Chiba University

\medskip
\begin{center}
{\bf \large Abstract}
\end{center}

\vspace{-0cm}

Estimating income distributions plays an important role in the measurement of inequality and poverty over space. 
The existing literature on income distributions predominantly focuses on estimating an income distribution for a country or a region separately and the simultaneous estimation of multiple income distributions has not been discussed in spite of its practical importance. 
In this work, we develop an effective method for the simultaneous estimation and inference for area-wise spatial income distributions  taking account of geographical information from grouped data. 
Based on the multinomial likelihood function for grouped data, we propose a spatial state-space model for area-wise parameters of  parametric income distributions. 
We provide an efficient Bayesian approach to estimation and inference for area-wise  latent parameters, which enables us to compute area-wise summary measures of income distributions such as  mean incomes and Gini indices, not only for sampled areas but also for areas without any samples thanks to the latent spatial state-space structure. 
The proposed method is demonstrated using the Japanese municipality-wise grouped income data. 
The simulation studies show the superiority of the proposed method to a crude conventional approach which estimates the income distributions separately.

\bigskip\noindent
{\bf Key words}: 
Grouped data; Income distribution; Markov Chain Monte Carlo; Pair-wise difference prior; Spatial smoothing

\newpage
\section{Introduction}
The estimation of income distributions has been widely recognized for playing an important role in the measurement of inequality, poverty and welfare over time and space. 
The comprehensive contents regarding the different parametric models for income distributions and and various estimation methods are available in \cite{KK2003} and \cite{Cho2008}. 
Information on income are generally obtained by household expenditure and income surveys in many countries. 
However, the availability of raw data is usually severely limited because of several problems such as the difficulty in data management and confidentiality of individual income data. 
Instead, the income data is typically provided as summary statistics (e.g. mean income and Gini index) and grouped data in the form of frequency distributions over some predefined income classes.
Based on such limited data, many statistical methods to estimate income distributions have been proposed so far \citep[e.g.][]{Cho2007, Cho2012, Haj2012, GH2015, KK2003, WP2007}.
A standard approach for estimating income distributions is to assume a parametric family of income distributions to approximate the true income distribution and estimate its unknown parameters based on the (limited) data.
There exists a wide variety of families of distributions available \citep[e.g.][]{HG2013, KK2003, Mc1984, MX1995, SM1976}. 
The existing approaches predominantly focus on the income distribution using the data only from a single area, e.g. country or state,  in a single period. 
Although there is some development in the direction of estimating a series of income distributions using time series data \citep{Nishino2012, Nishino2015}, there exists no work which estimates income distributions of multiple areas simultaneously taking spatial dependencies into account.

Estimating income distributions and associated inequality measures over multiple local areas such as municipalities and comparing them is particularly crucial to capture local economic status for local policy making. 
Even in this multiple income distribution scenario, the conventional methods would still separately estimate the area-wise income distributions based on area-wise grouped data. 
The separate estimation would be reasonable if the study is on a national scale such as \cite{Cho2007}. 
However, the conventional methods would encounter the following problems. 
One of the biggest problems is that the sample sizes can be very small in some local areas because of small populations and for such areas the estimates of the income distributions can be inaccurate and unstable. 
Furthermore, there are situations where a survey does not take sample from some of the areas. 
This is particularly the case for the grouped data obtained from Housing and Land Survey (HLS) of Japan in 2013 where 634 out of 1899 municipalities in Japan were not sampled.  
It is impossible to infer the income distributions for the non-sampled areas unless there is a statistical model that connects the sampled areas and non-sampled areas. 
Moreover, although the income distributions of the neighboring local areas would exhibit similar tendency, the conventional estimates would exhibit a large variation among the neighboring areas due to the aforementioned reason. 
It would be natural to take account of geographical information. 
As these issues cannot be addressed by simply using the separate estimation, a more sophisticated method is required.

In this paper, we propose a simultaneous estimation and inference method for area-wise spatial income distributions taking account of geographical relationship from grouped data.
Specifically, we assume parametric families of income distributions with different parameter values over different areas and introduce a state-space model with a spatial stochastic structure for the spatially varying parameters. 
Since the propose method takes account of geographical information through a latent stochastic structure, the estimation accuracy in areas with small sample sizes can be improved by borrowing strength from adjacent areas. 
Furthermore, income distributions in non-sampled areas  can be estimated using the information from adjacent areas, which would be a significant advantage over the conventional separate estimation method.
Also, the latent structure tends to produce similar parameter values for adjacent areas, so that the resulting estimated income distributions exhibit the same tendency with the improved interpretability.
We develop a Bayesian estimation approach to the proposed model and provide the posterior computation algorithm based on Markov Chain Monte Carlo (MCMC), which facilitates the estimation as well as inference not only for area-wise parameters but also important summary measures such as mean incomes and Gini indices.

This paper is organized as follows.
Section \ref{sec:method}, introduces the proposed method and provide Bayesian estimating procedures.
Section \ref{sec:app}, the proposed method is demonstrated to HLS data to estimate the income distributions across Japan and  confirms that it produces the spatially smoothed estimates of the area-wise means and Gini indices. 
In Section \ref{sec:sim}, we conduct simulation studies to compare the performance of the proposed and the separate estimation method.
Finally, some conclusion and discussion are given in Section \ref{sec:conc}.

\section{Estimation and inference for area-wise income distributions}\label{sec:method}

\subsection{Grouped income data and multinomial likelihood}
Suppose there are $m$ areas and we are interested in the income distributions $F_i(\cdot), i=1,\ldots,m$.
Instead of individual incomes, we can only observe the grouped data which comprise the number of individuals $c_{ik}$ in the $k$th income group for $k=1,\dots,N$ where $N$ is the number of  income groups. 
The income groups are defined as $(z_0, z_1)$, $(z_1, z_2),\ldots,(z_{N-1}, z_N)$ with $z_0=0$ and $z_N=\infty$.  
In this paper, we assume that all the income distributions $F_1(\cdot),\ldots,F_m(\cdot)$ have the same parametric form but they could have different parameter values.
Let $\eta_i$ be a vector of parameters that characterizes $F_i(\cdot)$, that is, $F_i(\cdot)=F(\cdot; \eta_i)$.
There exist several parametric distribution families for income models, e.g. the generalized beta distribution and its special and limiting cases  the beta-2, Singh-Maddala, Dagum, generalized gamma distribution and lognormal distribution.
Once the parametric distribution is specified, the summary measures such as the mean and variance as well as the inequality measures such as Lorenz curve and Gini index can be immediately obtained. 
In Table 1 in \cite{Haj2012}, the several parametric distributions and forms of the important measures are summarized.
The dimension of the area-wise parameter $\eta_i$ depends on the choice of the parametric distribution for $F_i(\cdot)$ and typically ranges from 2 to 4.
The standard approach to estimating $\eta_i$ from the grouped data in the $i$th area is based on the multinomial likelihood function which is proportional to
\begin{equation}\label{Like}
M_i(\eta_i)=\prod_{k=1}^N\left\{F(z_k;\eta_i)-F(z_{k-1};\eta_i)\right\}^{c_{ik}}.
\end{equation}
By maximizing the above function with respect to $\eta_i$, the maximum likelihood estimator of $\eta_i$ is obtained.

In order to carry out the simultaneous estimation and inference for the multiple $\eta_i$'s, we treat them as latent variables and introduce a stochastic latent structure for them.
It should be noted that the parameter spaces of each element of $\eta_i$ can vary depending on the choice of $F_i(\cdot)$.
Hence, we first assume that there exists a function $h_{\ell}$ such that $\eta_{i\ell}\ (\ell=1,\ldots,p)$ can be expressed as $\eta_{i\ell}=h_\ell(u_{i\ell})$ for some $u_{i\ell}\in \Re$. 
This assumption is quite reasonable since it is satisfied by all the distributions given in Table 1 of \cite{Haj2012}. 
For instance, when we use the log-normal distribution with the mean parameter $\eta_{i1}$ and variance parameter $\eta_{i2}$, a reasonable choice of $h_1$ and $h_2$ would be $h_1(x)=x$ and $h_2(x)=\exp(x)$.
Since $u_{i\ell}$'s are on the whole real line, we can readily introduce spatial correlation among $u_{1\ell},\ldots,u_{m\ell}$.

\subsection{Latent models for area-wise parameters}
In order to express the area-wise heterogeneity and spatial similarity (parameter values would be similar for adjacent areas), we introduce the latent model that defines a structure among area-wise parameters.
To this end, the stochastic structures for both the area-wise parameters themselves and differences of the parameters are introduced.
Specifically, we assume that the following prior for $(u_{1\ell},\ldots,u_{m\ell})$:
\begin{equation}\label{prior}
\pi(u_{1\ell},\ldots,u_{m\ell})
\propto C(\tau_\ell,\la_\ell)\exp\left\{-\frac{\tau_\ell}{2}\sum_{i=1}^m(u_{i\ell}-\mu_\ell)^2-\frac{\la_\ell}{2}\sum_{i<j}^m w_{ij}(u_{i\ell}-u_{j\ell})^2\right\},
\end{equation}
independently for $\ell=1,\ldots,p$ where $w_{ij}$ is equal to one if the area $j$ is  adjacent the area $i$ ($i\sim j$) and zero otherwise ($i\sim \hspace{-1em} \slash\ j$). 
The first part has a role to shrink $u_{i\ell}$ towards the grand mean $\mu_{\ell}$ and the second part shrinks the difference between $u_{i\ell}$ and $u_{j\ell}$ of two adjacent areas toward $0$.
The specification of the second part is included in the class of pair-wise difference priors \citep{Besag1995} whose form is the same as the well-known conditional autoregressive prior \citep{Besag1974, BK1995}. 
In (\ref{prior}), $\tau_\ell$ and $\la_\ell$ are precision parameters in the first and second parts, and $C(\tau_\ell,\la_\ell)$ is the normalizing constant relevant to the precision parameters, that is, $C(\tau_\ell,\la_\ell)=|Q(\tau_\ell,\lambda_\ell)|^{1/2}$ with $Q(\tau,\la)=\tau I_m+\la W_{\ast}$, where $W_{\ast}={\rm diag}(w_1,\ldots,w_m)-W$, $w_i$ is the number of adjacent areas and $W$ is an $m\times m$ matrix with the $(i,j)$th entry equal to $w_{ij}$.

The specification of the latent stochastic structure of the area-wise parameter $u_i$ ($\eta_i$) given in (\ref{prior}) combined with the multinomial likelihood functions (\ref{Like}) gives the state-space model for the grouped data.
The unknown structural parameters in the proposed model are the grand mean $\mu_\ell$ and precision parameters $\tau_\ell$ and $\la_\ell$ for $\ell=1,\dots,p$.
We develop an efficient Bayesian approach to estimate the area-wise parameter $\eta_i$ by assigning prior distributions for the unknown parameters.

\subsection{Posterior computation}\label{sec:post}
We consider the conditionally conjugate priors for the unknown parameters: $\mu_\ell\sim N(0,a_\mu^{-1})$, $\tau_\ell\sim \Gamma(b_\tau,c_\tau)$ and $\la_\ell\sim \Gamma(b_\la,c_\la)$ with $a_\mu,b_\tau,c_\tau,b_\la$ and $c_\la$ specified by the analyst.
In this paper, we set $a_\mu=10^{-6}$ and $b_\tau=c_\tau=b_\la=c_\la=1$ as default choice. 
The likelihood  associated with the joint distribution of $(c_{i1},\dots,c_{iN}), \ i=1,\dots,m$ and $(u_{1\ell},\dots,u_{m\ell}), \ \ell=1,\dots,p$ is given by 
\begin{equation*}
\left\{\prod_{\ell=1}^pC(\tau_{\ell},\la_{\ell})\right\}
\exp\Big\{\sum_{i=1}^m L_i(u_i)-\frac12\sum_{\ell=1}^p\sum_{i=1}^m\tau_\ell (u_{i\ell}-\mu_\ell)^2-\frac12\sum_{\ell=1}^p\sum_{i<j}^m \la_\ell w_{ij}(u_{i\ell}-u_{j\ell})^2\Big\},
\end{equation*} 
where $L_i(u_i)=\log M_i(h_1(u_{i1}),\ldots,h_p(u_{ip}))$ with $M_i$ defined in (\ref{Like}).
We now outline the posterior sampling algorithm for the proposed model. 
The details are presented later, but in brief, the algorithm proceeds by repeating the following four steps:
\begin{itemize}
\item[1.]
For $i=1,\ldots,m$, update the $p$-dimensional vector of the latent area-wise parameters $u_i$ using the independent Metropolis-Hastings (MH) algorithm. 
The detail is provided below.

\item[2.]
For $\ell=1,\ldots,p$, update the overall mean $\mu_{\ell}$ from the full conditional posterior $N(\mu_{\ast},\sigma_\mu^2)$ where $\mu_{\ast}=(m\tau_\ell+a_\ell)^{-1}\tau_\ell\sum_{i=1}^mu_{i\ell}$, $\sigma_\mu^2=(m\tau_\ell+a_\ell)^{-1}$.  

\item[3.]
For $\ell=1,\ldots,p$, update the precision parameters $(\tau_{\ell},\la_\ell)$ using the Metropolis adjusted Langevin algorithm whose details are provided below.
\end{itemize}

In Step~1, we need to sample from the full conditional distribution of $u_{i\ell}$ which is proportional to
\begin{equation}\label{full-u}
h(u_i)=\exp\Big\{L_i(u_i)-\frac12\sum_{\ell=1}^p\tau_{\ell}(u_{i\ell}-\mu_\ell)^2-\frac12\sum_{\ell=1}^p\sum_{j=1}^m\la_{\ell}w_{ij}(u_{i\ell}-u_{j\ell})^2\Big\}.
\end{equation}
This form is not in a standard form due to the complicated multinomial likelihood $\exp(L_i)$ of the form (\ref{Like}). 
In order to carry out independent MH algorithm efficiently, we construct a proposal distribution by approximating the full conditional distribution by a normal distribution.
To this end, $\exp(L_i)$ is approximated by the $p$-dimensional normal density function with the mean vector and precision matrix  set to the mode and negative Hessian matrix at the mode of $L_i(\cdot)$ denoted by $\ut_i$ and $P_i$, respectively. 
Then the full conditional distribution proportional to (\ref{full-u}) can be approximated by $\phi_p(\cdot;\alpha_u,\Gamma_u)$, where $\alpha_u=\Gamma_u(P_i\ut_i+T\mu+\Lambda\sum_{j=1}^mw_{ij}u_j)$ and $\Gamma_u=(P_i+T+w_i\Lambda)^{-1}$ with $T=\diag(\tau_1,\ldots,\tau_p)$, $\Lambda=\diag(\la_1,\ldots,\la_p)$ and $\mu=(\mu_1,\ldots,\mu_p)^t$.
A proposal $u_i^{\ast}$ is generated from this approximated full conditional distribution independent to the current value $u_i$ and is accepted  with the probability $\min\{1,h(u_i^{\ast})\phi_p(u_i;\alpha_u,\Gamma_u)/h(u_i)\phi_p(u_i^{\ast};\alpha_u,\Gamma_u)\}$.

In Step-3, with the current values $\tau_{\ell}^{\dagger}$ and $\la_{\ell}^{\dagger}$, the candidates of $\tau_{\ell}$ and $\la_{\ell}$ are generated from the following Langevin diffusion:
$$
(\tau_{\ell},\la_{\ell})=(\tau_{\ell}^{\dagger},\la_{\ell}^{\dagger})
-h\nabla U(\tau_{\ell}^{\dagger},\la_{\ell}^{\dagger})+\sqrt{2h}\ep_{\ell},
$$ 
where $\ep_{\ell}\sim N(0,I)$, and $U(\tau_{\ell},\la_{\ell})$ is the negative log-posterior distribution of $\tau_{\ell},\la_{\ell}$.
The partial derivatives of $U(\tau_{\ell},\la_{\ell})$ are obtained as 
\begin{align*}
&\frac{\partial}{\partial\tau_\ell}U(\tau_{\ell},\la_{\ell})
=\frac12{\rm tr}\{Q(\tau_{\ell},\la_{\ell})^{-1}\}+\frac12\sum_{i=1}^m(u_{i\ell}-\mu_\ell)^2-\log\pi(\tau_{\ell}),\\
&\frac{\partial}{\partial\tau_\ell}U(\tau_{\ell},\la_{\ell})
=\frac12{\rm tr}\{Q(\tau_{\ell},\la_{\ell})^{-1}W_{\ast}\}+\frac12\sum_{i<j}^m w_{ij}(u_{i\ell}-u_{j\ell})^2-\log\pi(\la_{\ell}),
\end{align*} 
where $\pi(\tau_{\ell})$ and $\pi(\la_{\ell})$ are prior distribution for $\tau_\ell$ and $\la_{\ell}$, respectively.
The proposed values $(\tau_{\ell}^{\ast},\la_{\ell}^{\ast})$ are accepted with the probability ${\rm min}(1,r_\ell)$ with 
$$
r_\ell=
\frac{h(\tau_\ell^{\ast},\la_\ell^{\ast})\phi((\tau_\ell^{\dagger},\la_\ell^{\dagger}),(\tau_\ell^{\ast},\la_\ell^{\ast})+h\nabla U(\tau_\ell^{\ast},\la_\ell^{\ast}),2hI_2)}{h(\tau_\ell^{\ast},\la_\ell^{\ast})\phi((\tau_\ell^{\ast},\la_\ell^{\ast}),(\tau_\ell^{\dagger},\la_\ell^{\dagger})+h\nabla U(\tau_\ell^{\dagger},\la_\ell^{\dagger}),2hI_2)},
$$
where $\phi(\cdot; \mu,\Sigma)$ is a multivariate normal density with the mean vector $\mu$ and variance-covariance matrix $\Sigma$, and $h(\tau_\ell,\la_\ell)$ includes the relevant terms of the full conditional distribution given by
$$
h(\tau_\ell,\la_\ell)
\propto C(\tau_\ell,\la_\ell)\exp\left\{-\frac{\tau_\ell}{2}\sum_{i=1}^m(u_{i\ell}-\mu_\ell)^2-\frac{\la_\ell}{2}\sum_{i<j}^m w_{ij}(u_{i\ell}-u_{j\ell})^2\right\}\pi(\tau_{\ell})\pi(\la_{\ell}). 
$$

Based on the posterior samples of $u_i$, the estimates of the area-wise income distributions and important summary measures such as area-wise means and Gini indices can be computed. 
The uncertainty quantification (e.g. the posterior standard error and credible interval) of the estimates can be easily carried out based on the posterior samples.
Moreover, we can easily incorporate the sampling steps of  the areal parameters for non-sampled areas. 
Let $(s)$ denotes the index of the non-sampled area, so that $u_{(s)}$ represents the parameters of the income distribution in the area $(s)$. 
Since there is no data in the area $(s)$, $L_i(u_i)$ in (\ref{full-u}) does not exist and the full conditional distribution of $u_{(s)\ell}$ is simply $N(\mu_{(s)},\sigma_{(s)}^2)$ with $\mu_{(s)}=\sigma_{(s)}^2(\tau_\ell\mu_\ell+\lambda_\ell\sum_{j=1}^mw_{(s)j}u_{j\ell})$, $\sigma_{(s)}^2=(\tau_\ell+w_{(s)}\lambda_\ell)^{-1}$, $w_{(s)j}=1$ if $(s)\sim j$ and 0 otherwise, and $w_{(s)}=\sum_{j=1}^m w_{(s)j}$, the number of areas adjacent to the area $(s)$.

\subsection{Laplace-type distributions for pair-wise difference}
While the pair-wise difference prior in the form given in  (\ref{prior}) is widely used and can produce spatially smoothed estimates of parameters, we may use other forms of pair-wise difference to achieve the following.
When the income distributions of the two adjacent areas are quite similar with the similar values of $u_i$ and $u_j$ with $i\sim j$, setting  $u_i=u_j$ would be more interpretable. 
Moreover, when the difference between $u_i$ and $u_j$ for $i\sim j$ is sufficiently large, the pair-wise difference prior (\ref{prior}) based on the normal distribution might produce over-smoothed estimates of these parameters. 
In order to address these issues, we introduce the following latent structure:
\begin{equation}
\begin{split}\label{Lap}
\pi&(u_{1},\ldots,u_{m})\\
&= C_{\ast}(\tau,\la)
\exp\left[-\frac{1}{2}\sum_{i=1}^m\sum_{\ell=1}^p\tau_\ell(u_{i\ell}-\mu_\ell)^2 - \sum_{ \{ i\sim j, \ i<j \} } \la \left\{\sum_{\ell=1}^p(u_{i\ell}-u_{j\ell})^2\right\}^{1/2}\right],
\end{split}
\end{equation}
where $C_{\ast}(\tau,\la)$ is the normalizing constant.
Note that the second part will be included in the general pair-wise difference prior given in \cite{Besag1995} when $p=1$.
For general $p$, some similar forms have appeared in Bayesian group Lasso \citep{XG2015} and Bayesian elastic net \citep{LL2010}.
We use the same gamma prior for $\lambda$: $\la\sim \Gamma(b_\la,c_\la)$.
Using the identity,
$$
\exp(-\la z^{1/2}) = {2 \over \la} \int_0^\infty {1 \over \sqrt{2\pi s}}e^{-z / (2s)} {\la^2 \over 2}e^{-\la^2 s / 2}\mathrm{d}s, \quad \la > 0, \ z > 0,
$$
which is known as the normal mixture representation of the Laplace distribution \citep[e.g.][]{PC2008}, the distribution (\ref{Lap}) has the following hierarchical expression:
\begin{equation*}
\begin{split}
&\pi(u_{1},\ldots,u_{m}|S)\propto 
\exp\left[-\frac{1}{2}\sum_{i=1}^m\sum_{\ell=1}^p\tau_\ell(u_{i\ell}-\mu_\ell)^2
-\frac{1}{2}\sum_{ \{ i\sim j,\ i<j \} }\sum_{\ell=1}^p\frac{(u_{i\ell}-u_{j\ell})^2}{s_{ij}}\right],
\end{split}
\end{equation*}
where $S$ is the set of $s_{ij}$'s $(i\sim j, i<j)$ and the distribution of $S$ is set such as the joint distribution of $u_1,\dots,u_m$ and $S$ is the following form:
\begin{equation*}
\begin{split}
&\pi(u_1,\dots,u_m,S) \\
=& \ C_*(\tau,\la) \exp\left\{ -{1 \over 2}\sum_{i=1}^m\sum_{\ell=1}^p \tau_\ell(u_{i\ell} - \mu_\ell)^2 \right\} \left( {2 \over \la} \right)^\delta (2\pi)^{-\delta/2} \\
&\times \exp\left\{ -{1 \over 2}\sum_{ \{ i\sim j,\ i<j \} } \sum_{\ell=1}^p {(u_{il} - u_{jl})^2 \over s_{ij}} \right\} \times \prod_{ \{ i\sim j, \ i<j \} } s_{ij}^{-1/2}{\la^2 \over 2}\exp\left( -{\la^2 \over 2}s_{ij} \right),
\end{split}
\end{equation*}
for $\de = \sum_{i<j} w_{ij}$.
Since the joint distribution of $(u_{1\ell},\ldots,u_{m\ell})$ is the multivariate normal given $s_{ij}$'s, which makes the posterior computation much easier, that is, sampling from the full conditional posterior of $u_{i\ell}$ can be carried out in almost the same way as used in Section \ref{sec:post}.
Based on the above hierarchical expression, $C_{\ast}(\tau,\la)$ can be expressed as 
\begin{align*}
C_{\ast}&(\tau,\la)^{-1}\\
&\propto
\la^{-\delta}\iint \exp\left\{-\frac{1}{2}\sum_{i=1}^m\sum_{\ell=1}^p\tau_\ell(u_{i\ell}-\mu_\ell)^2
-\frac{1}{2}\sum_{ \{ i\sim j, \ i<j \} }\sum_{\ell=1}^p\frac{(u_{i\ell}-u_{j\ell})^2}{s_{ij}}\right\}
{\rm d}u \\
&\quad \times \prod_{ \{ i\sim j,\ i<j \} }^ms_{ij}^{-1/2}\pi(s_{ij}){\rm d}S\\
&\propto
\la^{-\delta}\int\prod_{\ell=1}^p|Q_{\ast}(\tau_\ell,S)|^{-1/2}\prod_{ \{ i\sim j, \ i<j \} } s_{ij}^{-1/2}\pi(s_{ij}){\rm d}S,
\end{align*}
where $\pi(s_{ij})$ is the density function of $\mathrm{Exp}(\la^2/2)$, that is, $\pi(s_{ij}) = \la^2/2 \exp(-\la^2s_{ij}/2)$, $Q_{\ast}(\tau_\ell,S)=\tau_\ell I_m+{\rm diag}(\st_1,\ldots,\st_m)-\widetilde{S}$, $\widetilde{S}$ is an $m\times m$ matrix whose $(i,j)$-element is $\st_{ij}$ for
$$
\st_{ij} =
\begin{cases}
1 / s_{ij} & i \sim j, \ i<j, \\
1 / s_{ji} & i \sim j, \ i>j, \\
0 & \mathrm{otherwise},
\end{cases}
$$
and $\st_i=\sum_{j=1}^m \st_{ij}$, row sum of $\widetilde{S}$.
The integral can be evaluated via simple Monte Carlo method generating a number of random samples of $s_{ij}$ from ${\rm Exp}(\la^2/2)$.

The full conditional distribution of $u_i$ using the hierarchy is given by
\begin{equation}\label{full-u2}
h_L(u_i)\propto\exp\Big\{L_i(u_i)-\frac12\sum_{\ell=1}^p\tau_{\ell}(u_{i\ell}-\mu_\ell)^2-\frac12\sum_{\ell=1}^p\sum_{j=1}^m \st_{ij} (u_{i\ell}-u_{j\ell})^2\Big\},
\end{equation}
so that we can use the similar sampling strategy used in the previous section. 
The sampling algorithm for the Laplace-type prior is given as follows.  
\begin{itemize}

\item[1.]
Generate samples of $\Phi=(\tau_1,\ldots,\tau_p,\la)$ via random-walk MH algorithm where a candidate value $\Phi^{\ast}$ of $\Phi$ is generated from $\Phi^{\dagger}+bN(0,I_{p+1})$ with the current value $\Phi^{\dagger}$ and step size $b$ and is accepted with the probability ${\rm min}(1,r_L)$ where $r_L=h(\Phi^{\ast})/h(\Phi^{\dagger})$ and $h(\Phi)$ is the full conditional distribution given by
$$
h(\Phi)\propto C_{\ast}(\tau,\la)\exp\left[-\frac{1}{2}\sum_{\ell=1}^p\sum_{i=1}^m\tau_\ell(u_{i\ell}-\mu_\ell)^2-\la\sum_{ \{ i\sim j,\ i<j \} } \left\{\sum_{\ell=1}^p(u_{i\ell}-u_{j\ell})^2\right\}^{1/2}\right]\pi(\Phi)
$$
and $\pi(\Phi)$ is the prior distribution of $\Phi$.

\item[2.]
For $(i,j)\in \{1,\ldots,m\}^2$ with $i\sim j$ and $i<j$, update $s_{ij}^{-1}$ from the full conditional distribution $\text{IG}(\mu',\la^2)$ where $\mu'=\sqrt{\la^2/\sum_{\ell=1}^p(u_{i\ell}-u_{j\ell})^2}$ and $\text{IG}(a,b)$ denotes the inverse-Gaussian distribution \citep{CF1989} with the density given by
$$
f(x)=\sqrt{\frac{b}{2\pi}}x^{-3/2}\exp\left\{-\frac{b(x-a)^2}{2a^2x}\right\}, \ \ \ x>0.
$$

\item[3.]
For $i=1,\ldots,m$, update the $p$-dimensional vector of the latent area-wise parameters $u_i$ using the independent MH algorithm.
The proposal $u_i^{\ast}$ is first generated from $N_p(\alpha_u^L,\Gamma_u^L)$ where $\alpha_u^L=\Gamma_u^L(P_i\ut_i+T\mu+\sum_{j=1}^m \st_{ij} u_j)$ and $\Gamma_u^L=(P_i+T+I_p\sum_{j=1}^m \st_{ij} )^{-1}$ and is accepted  with the probability $\min\left\{1,h_L(u_i^{\ast})\phi_p(u_i;\alpha_u^L,\Gamma_u^L) \right.$ $\left. / h_L(u_i)\phi_p(u_i^{\ast};\alpha_u^L,\Gamma_u^L)\right\}$ for the current value $u_i$.

\item[4.]
Generate samples of $\mu_{\ell}$ in the same way as Section \ref{sec:post}.

\end{itemize}

For the non-sample area $(s)$, again, the posterior samples of parameters $u_{(s)}$ can be generated.
For the current value $u_{(s)}$, first generate $s_{(s)j}$ for $j\sim (s)$ is generated in the same way as Step 2 and define $\st_{(s)j}$ as $\st_{(s)j} = s_{(s)j}^{-1}$ for $j \sim (s)$ and $\st_{(s)j} = 0$ otherwise.
Then  $u_{(s)}$ is updated by sampling from $N_p(\alpha_{(s)}^L,\Gamma_{(s)}^L)$ where $\alpha_{(s)}^L=\Gamma_{(s)}^L(T\mu+\sum_{j=1}^m \st_{(s)j} u_j)$ and $\Gamma_{(s)}^L=(T+I_p\sum_{j=1}^m \st_{(s)j} )^{-1}$.

\section{Simulation studies}\label{sec:sim}
Here the performance of the proposed methods is investigated via the Monte Carlo simulations.
In this study, we use the log-normal distribution as the underlying income distributions where the true parameters vary over areas.
The number of areas is set to $m=200$. 
To generate a synthetic dataset, we first generated the latitude and longitude of the areas from the uniform distribution on $(-1,1)$. 
Based on the location $s_i=(s_{i1},s_{i2})\in (-1,1)^2, \ i=1,\ldots,m$, we constructed an adjacent matrix whose diagonal elements are $0$ and $(i,j)$-element ($i\neq j$) is 1 if $\|s_i-s_j\|<0.15$ and 0 otherwise.
The average number of adjacent areas is about $3.3$, which is similar to that of the adjacent matrix of the Japanese municipalities used in Section \ref{sec:app}.
In the log-normal distribution, the mean and variance permeates  are defined as $\mu_i=u_{i1}$ and $\sigma_i^2=\exp(u_{i2})$, respectively, where $u_i=(u_{i1},u_{i2})$ is the (transformed) area-wise parameter on $\Re^2$.
We consider a situation where the area-wise parameters relatively smoothly change over the locations as observed in the results in Section \ref{sec:app}. 
Specifically, the following three scenarios for $u_i$ are used :
\begin{align*}
&\text{(A)} \ u_{i1}=0.1+s_{i1}^2+s_{i2}^2, \ \ \ u_{i2}=0.2s_{i1}+0.2s_{i2},\\
&\text{(B)} \ u_{i1}=0.1+s_{i1}^2+s_{i2}^2-0.3I(s_i\in G_1)+0.3I(s_i\in G_2), \ \ \ 
u_{i2}=0.2s_{i1}+0.2s_{i2},\\
&\text{(C)} \ (u_{i1},u_{i2})=(0.3,0.1)I(s_{i1}>0,s_{i2}>0)+(0.5,0.2)I(s_{i1}>0,s_{i2}<0)\\
& \hspace{3cm} +(0.7,0.3)I(s_{i1}<0,s_{i2}>0)+(1.0,0.4)I(s_{i1}<0,s_{i2}<0),
\end{align*}
where $G_1=\{x\in (-1,1)^2\ |\ \|x-(1/2,0)\|<0.3\}$ and $G_2=\{x\in (-1,1)^2\ |\  \|x+(1/2,0)\|<0.3\}$.
In Scenario (A), both mean and variance parameters change smoothly over areas, whereas in Scenario (B) the mean parameter has two discontinuous regions in which the mean parameter can be  smaller or larger by 0.3. 
In Scenario (C), on the other hand, the areas are divided into four groups  and the areas in the same groups have the same parameter values. 
For each area $i$, we generated $n_i$ samples from the log-normal distribution with the mean $\mu_i$ and variance $\sigma_i^2$ where $n_i$ was randomly generated from the discrete uniform distribution on $(50,300)$. 
Then the grouped data is constructed with the boundaries $(z_1,\ldots,z_6)=(2,4,6,8,10,15)$.

The proposed pair-wise difference (PWD) and pair-wise Laplace (PWL) priors are applied to the simulated dataset.  
The default values of the hyperparameters are used for both methods and we generated 2000 posterior samples of $u_i$'s after discarding 500 samples as a burn-in period. 
The posterior means of $u_i$'s are computed as the point estimates of $u_i$'s and $95\%$ credible intervals of $u_i$ are constructed based on the posterior samples.
For comparison, we also considered the area-wise maximum likelihood (AML) method which estimates the area-wise parameters by separately maximizing the multinomial likelihood as a crude alternative for estimating $u_i$.

\begin{figure}[!htb]
\centering
\includegraphics[width=15cm,clip]{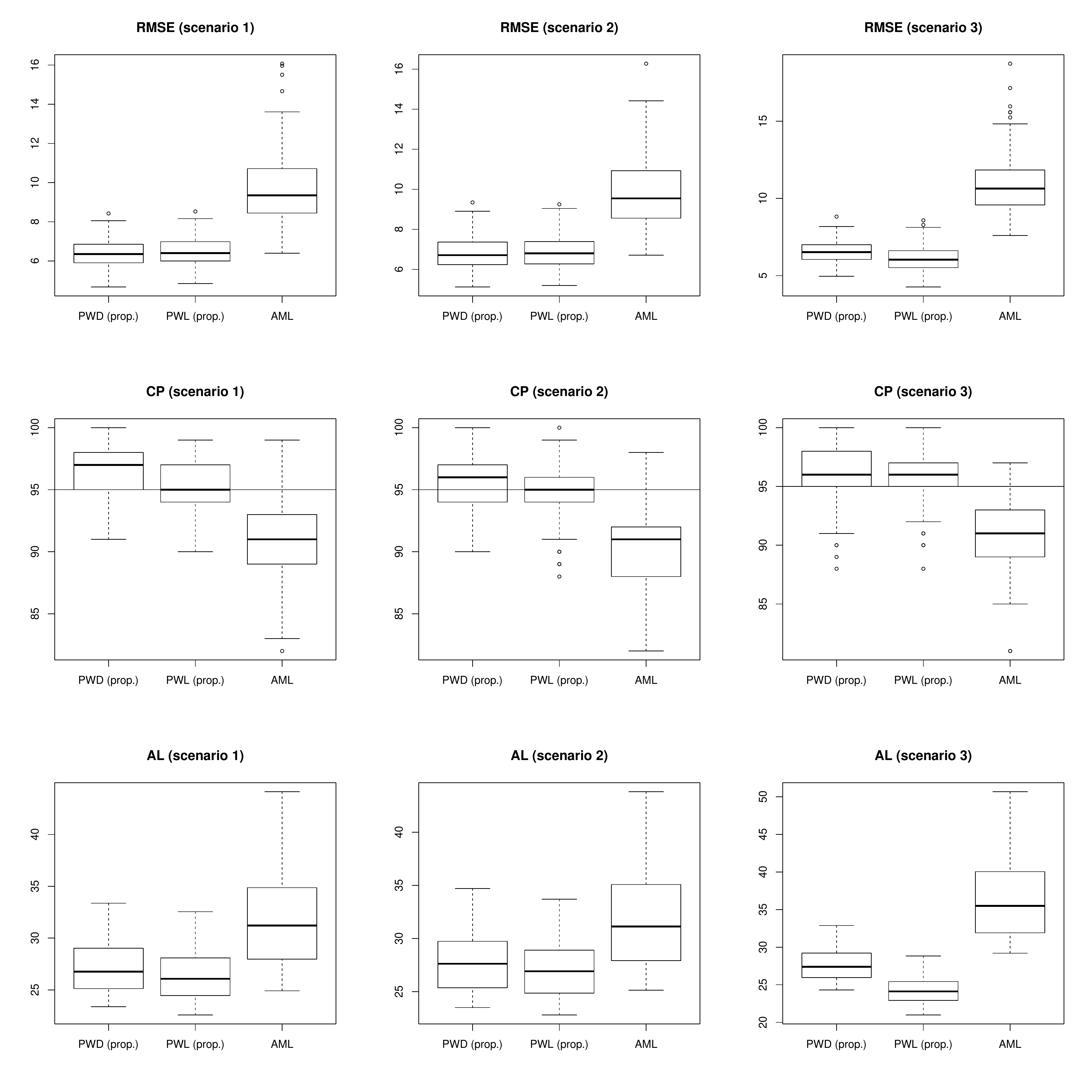}
\caption{
Boxplots of root squared mean squared errors (RMSE) of point estimates, and empirical coverage probabilities (CP) and average lengths (AL) of $95\%$ credible/confidence intervals of $u_{i1}$.
All values area multiplied by 100.
\label{fig:sim1}
}
\end{figure}

\begin{figure}[!htb]
\centering
\includegraphics[width=15cm,clip]{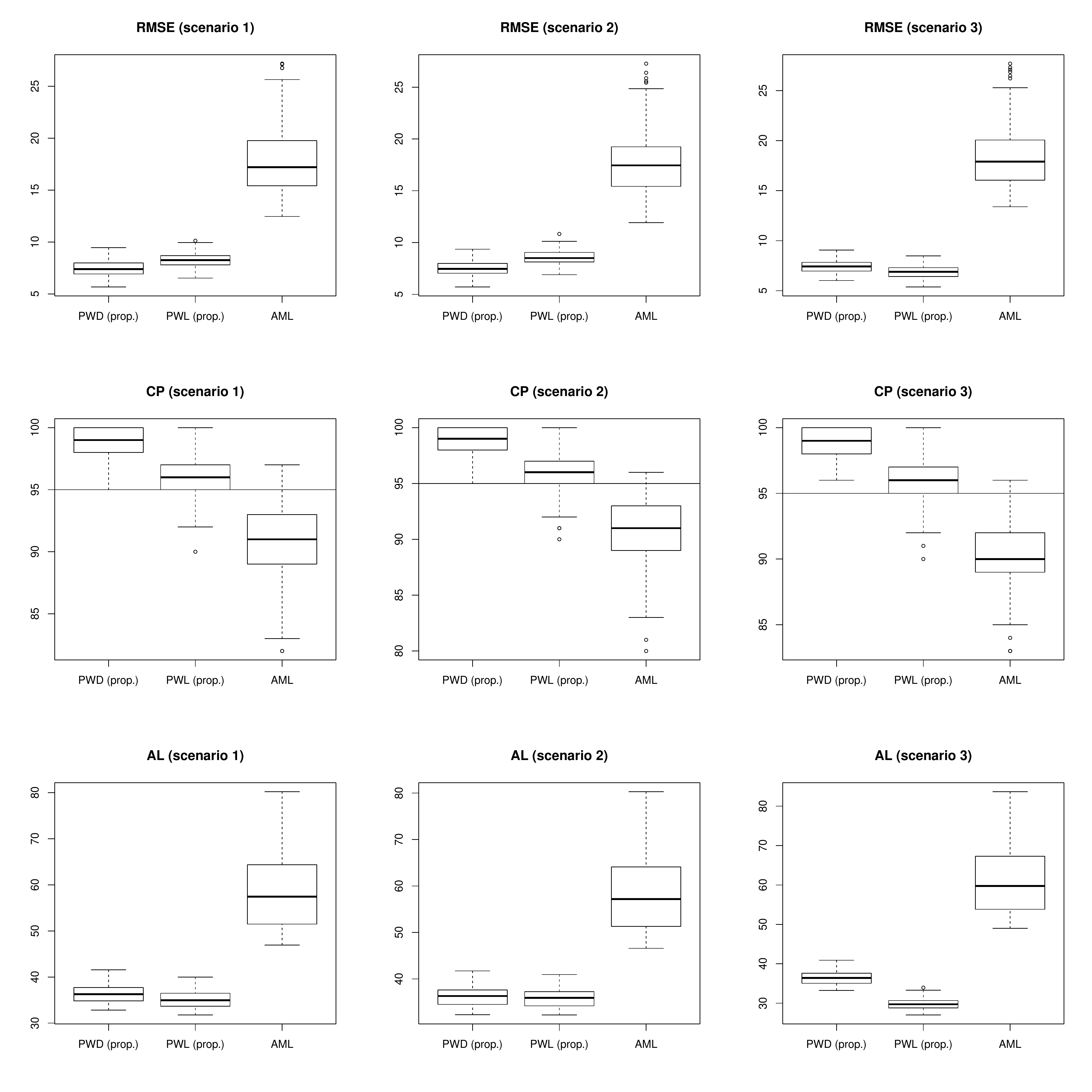}
\caption{
Boxplots of root squared mean squared errors (RMSE) of point estimates, and empirical coverage probabilities (CP) and average lengths (AL) of $95\%$ credible/confidence intervals of $u_{i1}$.
All values area multiplied by 100.
\label{fig:sim2}
}
\end{figure}

To evaluate the performance in the point estimation and interval estimation, we calculated the mean squared errors $\text{MSE}_{ik}=R^{-1}\sum_{r=1}^R(\widehat{u}_{ik}^{(r)}-u_{ik}^{(r)})^2$, coverage probability $\text{CP}_{ik}=R^{-1}\sum_{r=1}^RI(u_{ik}^{(r)}\in {\rm CI}_{ik}^{(r)})$, and average length of the credible intervals $\text{AL}_{ik}=R^{-1}\sum_{r=1}^R|{\rm CI}_{ik}^{(r)}|$ with $k=1,2$ and $R=100$ where $\widehat{u}_{ik}^{(r)}$ is a point estimate and ${\rm CI}_{ik}^{(r)}$ is a $95\%$ credible interval in the $r$th iteration.
Figures \ref{fig:sim1} and \ref{fig:sim2} report the boxplots of $\{\text{MSE}_{1k},\ldots,\text{MSE}_{mk}\}$, $\{\text{CP}_{1k},\ldots,\text{CP}_{mk}\}$ and $\{\text{AL}_{1k},\ldots,\text{AL}_{mk}\}$ for $k=1, 2$, respectively.
From the reported MSE, the proposed PWD and PWL clearly outperform the crude AML in terms of point estimation.
Moreover, both PWD and PWL produced some reasonable credible intervals whose coverage probabilities are approximately or larger than the nominal level of $95\%$ whereas  AML produced inefficient confidence intervals that tend to be wider with short coverage. 
Comparing PWD and PWL in terms of MSE, PWD performs better than PWL in both Scenario (A) while the opposite result was obtained in Scenario (C).
Note that the whole regions are divided into the finite number of groups in Scenario (C).
In such a case, PWL method seems to work better than PWD, which is consistent to the motivation of using PWL.
In terms of interval estimation, the coverage probabilities of PWL are generally close to the nominal level and the average lengths of PWL are smaller than PWD in all cases, which may reflect the efficiency of the Laplace formulation in PWL. 
The result would imply that PWL is more efficient than PWD in terms of interval estimation.

\section{Application to Japanese municipality-wise grouped income data}\label{sec:app}
The proposed method is demonstrated using the Japanese municipality-wise grouped data where we are interested in the spatial distribution of the local economic status of 1897 municipalities in Japan. 
The grouped income data is obtained from Housing and Land Survey (HLS) of Japan in 2013. 
Our dataset contains information on the number of working households that fall in the $9$ income class in 1263 municipalities out of 1897 municipalities. 
Note that the remaining 634 municipalities are not sampled in this data. 
The group boundaries are given by $(z_1,\ldots,z_8)=(1,2,3,4,5,7,10,15)$ (million yen). 
Since the exact sample size in each area is not known, the present study assumes that $1\%$ of populations are sampled, which could be a conservative value since most survey have larger sample sizes. 
See \cite{KawaKoba2019} for more details of the HLS data.

\begin{figure}[!htb]
\centering
\includegraphics[width=14cm]{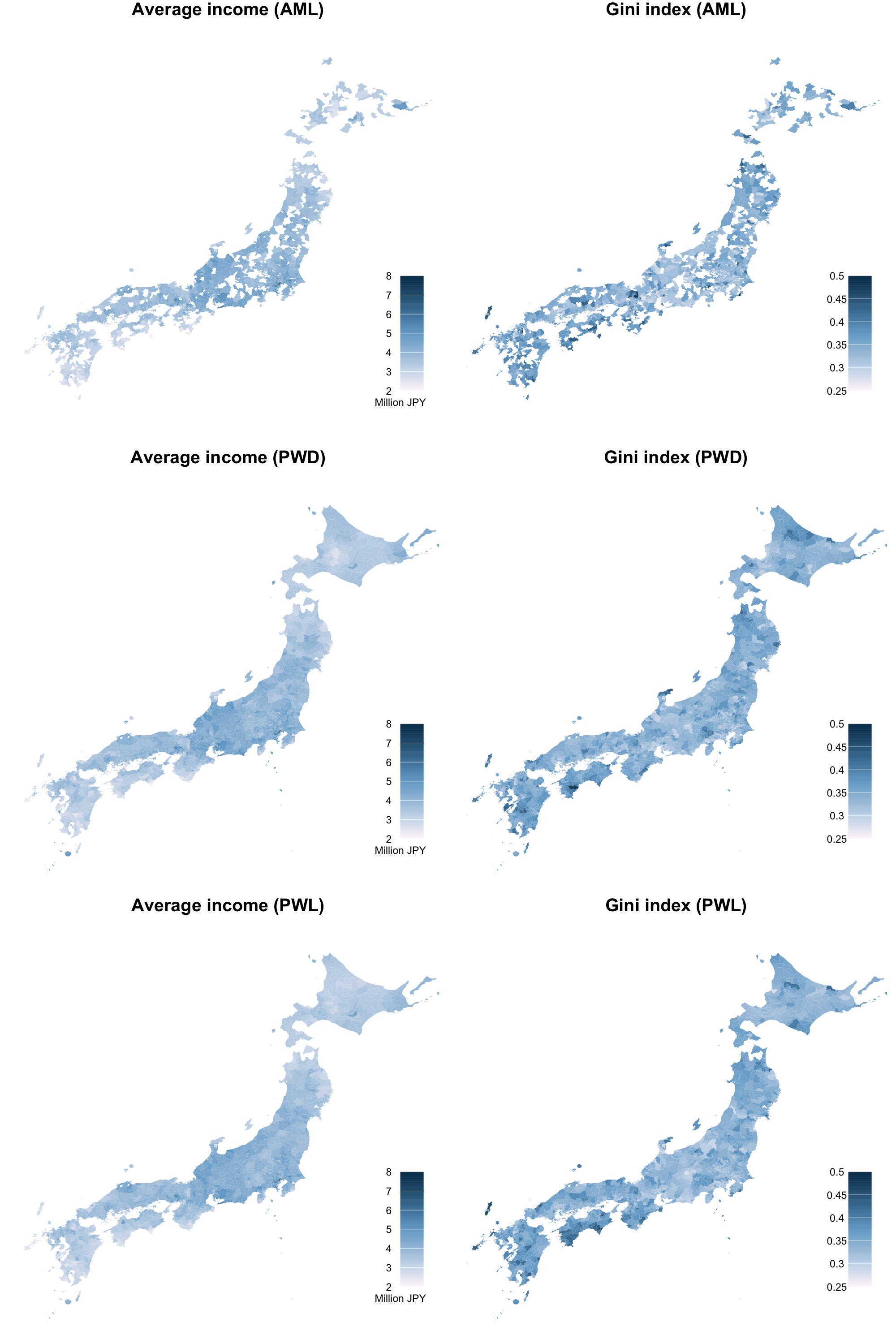}
\caption{
Estimates of area-wise mean and Gini index based on area-wise maximum likelihood (AML), pair-wise difference prior (PWD) and pair-wise Laplace-type prior (PWL).
\label{fig:Japan-est}
}
\end{figure}

First the AML method is applied using the log-normal distributions, LN$(\mu_i,\sigma_i^2), \ i=1,\ldots,m(=1263)$.
Then the average income, $\exp(\mu_i+\sigma_i^2/2)$, and Gini index, $2\Phi(\sigma_i/\sqrt{2})-1$ in each area are computed  based on the maximum likelihood estimates of $\mu_i$ and $\sigma_i^2$.
The results are shown in Figure~\ref{fig:Japan-est} where the estimated values under AML are highly variable over the areas.
Since some areas with small populations have very small sample sizes, the maximum likelihood estimates based only on the area-specific grouped data are unstable and inaccurate. 
Moreover, there would be no reasonable way for the AML method to estimate the parameters in the non-sampled areas, hence the practicality of the AML method is clearly limited.

In order to improve the estimation accuracy as well as to estimate the means and Gini indices in both sampled and non-sampled municipalities, we apply the six proposed methods based on the combinations of the popular three parametric income distributions, log-normal (LN), Singh-Madala (SM) and Dagum (DG) distributions, and the two different latent distributions, PWD and PWL priors.
Note that the density function of the SM and DG distributions are given by
\begin{align*}
&\text{SM}: \ \ f(x;a,b,c_1)=\frac{ac_1x^{a-1}}{b^a\{1+(x/b)^a\}^{1+c_1}},  \\ 
&\text{DG}: \ \ f(x;a,b,c_2)=\frac{ac_2x^{ac_2-1}}{b^{ac_2}\{1+(x/b)^a\}^{1+c_2}},
\end{align*}
where $a, c_1$ and $c_2$ are shape parameters and $b$ is a scale parameter.
We used $h(x)=\exp(x)$ for the parameter transformation in the proposed model.
With the default choices of the hyperparameters, we generated 5000 posterior samples of the area-wise parameters and structural parameters after discarding the first 500 samples.

The models are compared based on the following quantities mimicking posterior predictive loss \citep{GG1998}: 
$$
\text{PPL}({\cal M})=\frac1m\sum_{i=1}^m\sum_{k=1}^NV^{{\cal M}}_{ik}+\frac1{m+1}\sum_{i=1}^m\sum_{k=1}^N(c_{ik}-E^{{\cal M}}_{ik})^2,
$$
where $c_{ik}$ is the number of households belong to the $k$th group in the $i$th area, $E^{{\cal M}}_{ik}$ and $V^{{\cal M}}_{ik}$ respectively are the mean and variance of the posterior predictive distribution for $c_{ik}$ under model ${\cal M}$.
Table~\ref{tab:PPL} presents the values of PPL for the six models. 
It is observed that the SM distribution would be preferable choice among the three candidate distributions. 
The SM distribution is known to provide good fit to income data in many countries especially where the income distributions are more characterized by the upper tails than the lower tails \citep{KK2003, Kak2019}. 
Our result in which the SM distributions is supported by the data agrees with the past findings \citep{Mc2008, Atoda1988}. 
On the other hand, the DG distribution which has more rich characteristics in the lower tail may not be suitable for the Japanese income data. 
Comparing the two latent structures, the values of PWD and PWL are relatively comparable. 
The lognormal distribution is a simple income model, but it is known to provide poor fit in many cases due to its limitation in flexibility. 
In what follows, we will focus on the results of the proposed method based on the SM distribution in what follows.

We reported the posterior mean and $95\%$ credible intervals of the structural parameters in Table \ref{tab:para}, which shows that the results for the grand means of area-wise parameters are almost the same between PWD and PWL.
Comparing the values of $\lambda_\ell$'s and $\tau_\ell$'s, there would be some spatial correlations among municipalities as we expected.

\begin{figure}[!htb]
\centering
\includegraphics[width=15cm,clip]{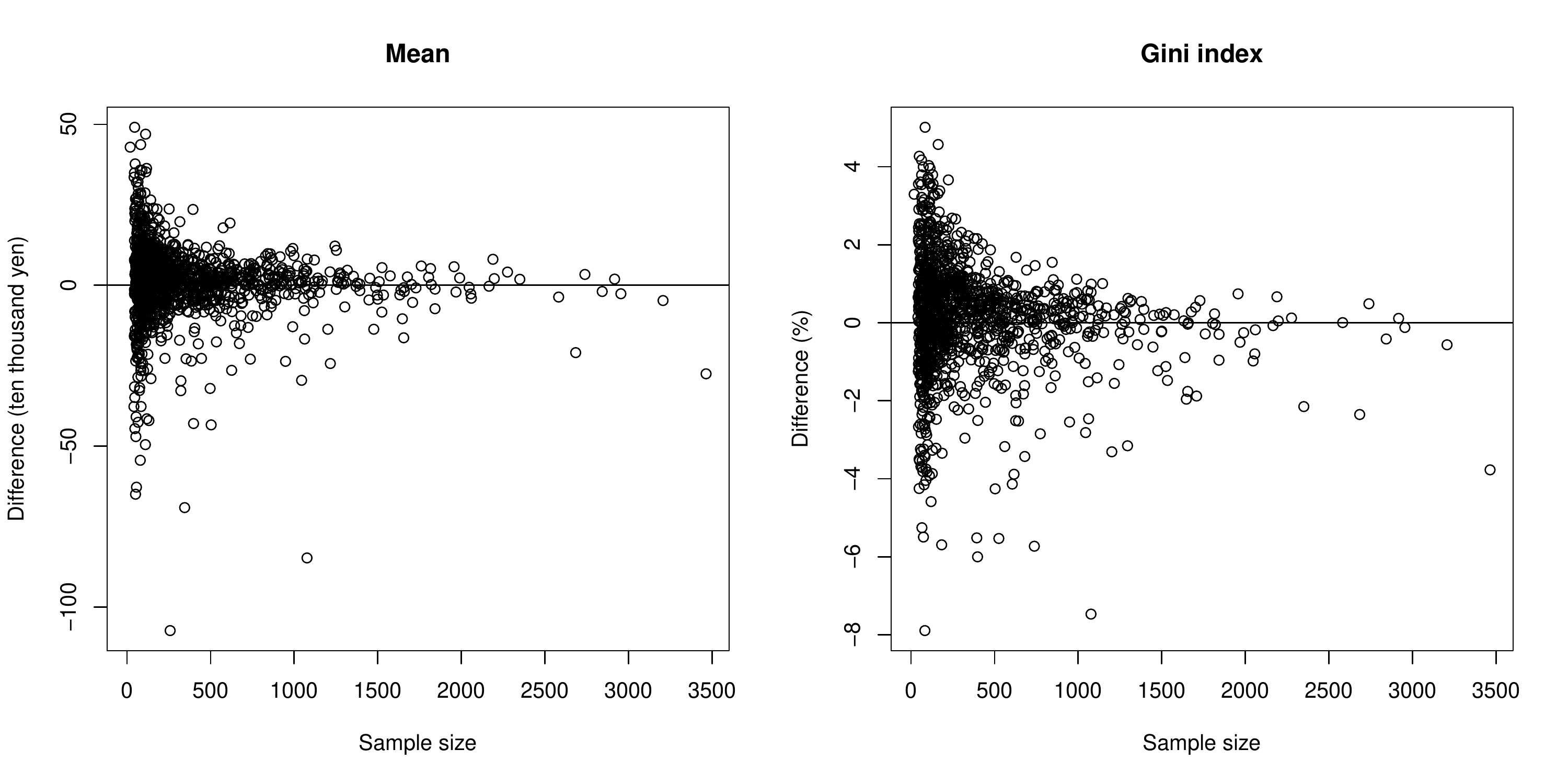}
\caption{
Differences of estimates between SM-PWD and SM-AML.
\label{fig:Japan-dif}
}
\end{figure}

\begin{figure}[!htb]
\centering
\includegraphics[width=15cm,clip]{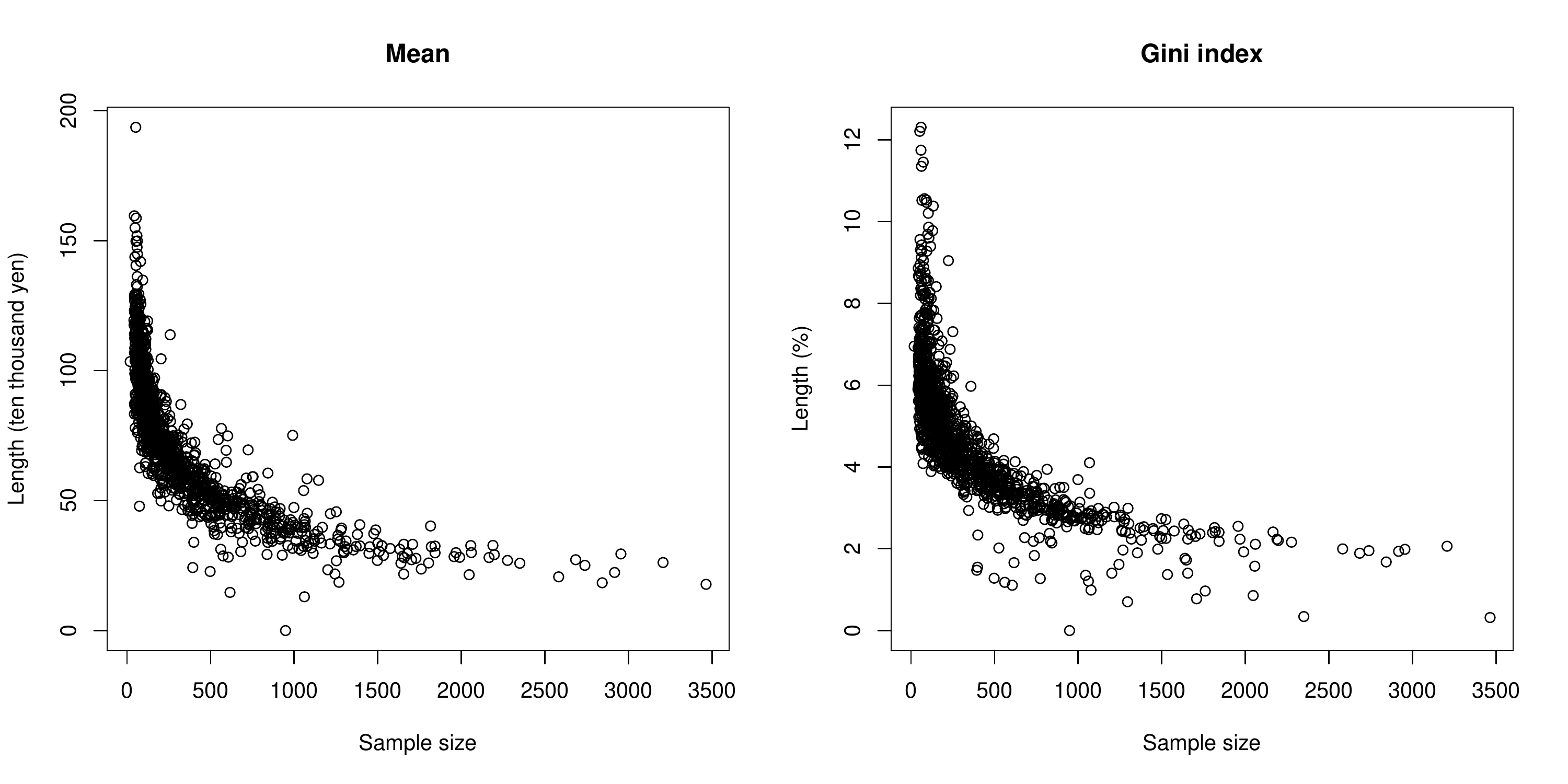}
\caption{
Lengths of 95\% credible intervals under SM-PWD.
\label{fig:Japan-CI}
}
\end{figure}

\begin{figure}[!htb]
\centering
\includegraphics[width=14cm,clip]{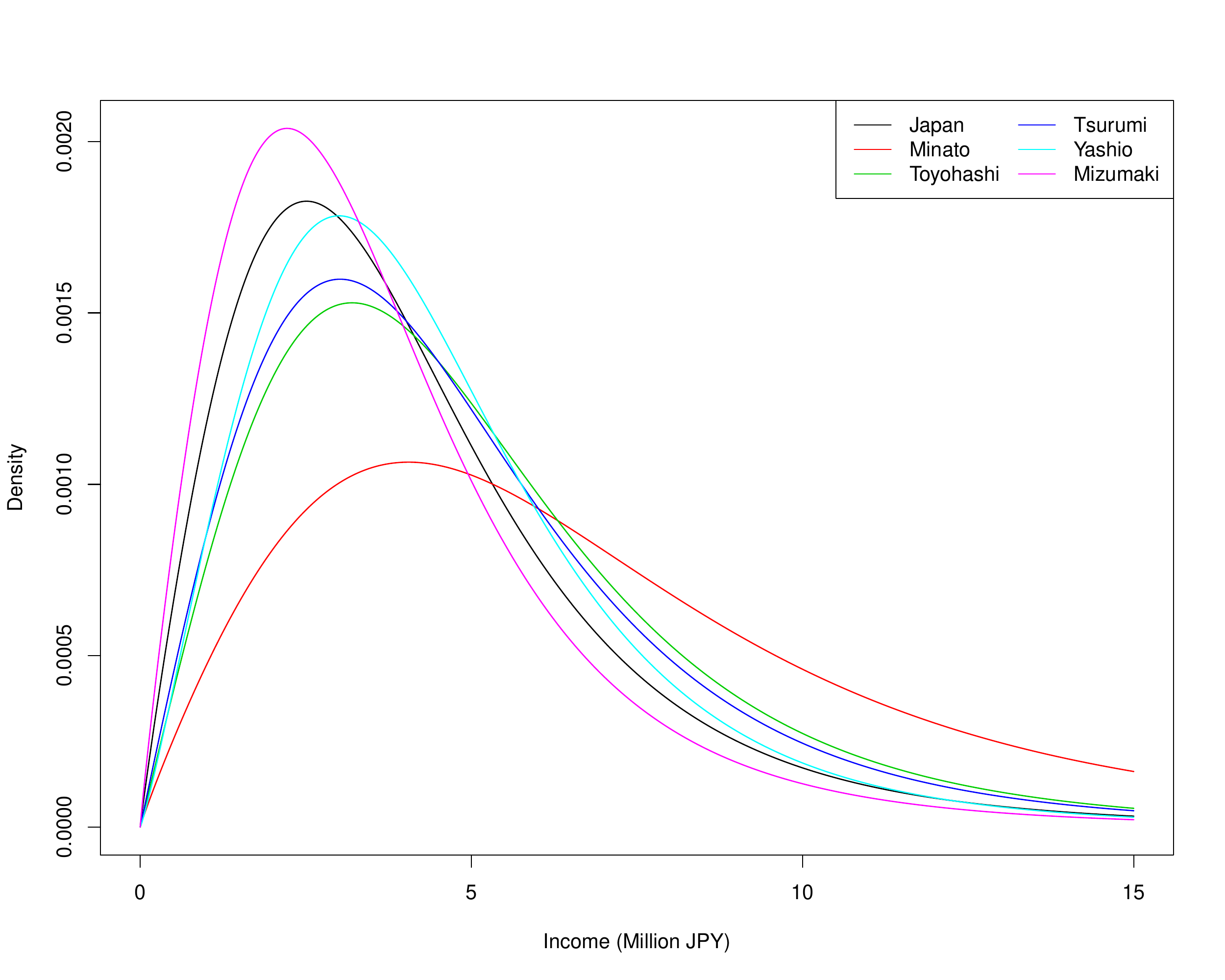}
\caption{
Estimated area-specific income distributions of the five selected areas and national income distribution under SM-PWD.
\label{fig:Japan-dens}
}
\end{figure}

Based on the posterior samples of area-wise parameters, the posterior means of the area-wise means and Gini indices in the sampled areas are computed.
Figure \ref{fig:Japan-dif} shows the differences in the estimates of the average incomes and Gini indices between  the proposed PWD and AML methods against the sample sizes in the sampled areas. 
It is observed that the proposed method produces almost the same estimates as AML in areas with large samples while a difference of estimates gets larger according to the sample size. 
The figure also indicates that in the areas with the small and moderate sample sizes, AML tends to produce the estimates slightly smaller than those under the proposed SM-PWD. 
In order to see the uncertainty of the estimates based on the proposed method, we calculated the 95\% credible intervals of the average incomes and Gini indices and the lengths of the intervals. 
Figure \ref{fig:Japan-CI}  confirms that the proposed method produces the credible intervals with the reasonable lengths in all  sampled areas. 
The figure also shows that the intervals becomes wider (uncertainty gets larger) as the sample sizes decrease. 
We next computed the area-specific density functions under SM-PWD using the posterior mean of the area-specific parameters of the SM distribution.
We selected 5 areas whose average incomes are ranked at the first, 100th, 200th, 500th and 1000th in descending order. 
The density of the national income distribution is also computed using the global parameters.
The six estimated densities are presented in Figure~\ref{fig:Japan-dens}. 
It is confirmed that the estimated area-wise income distributions vary considerably over the areas and are significantly different from the national income distribution in some areas.
Especially, the estimated income distribution in Minato area, which is located in  central Tokyo and has the largest estimated average income, exhibits much heavier tail than those of the other areas such as Mizumaki area in northern Kyushu whose estimated average income is the lowest among the five areas.

We now focus on the average incomes and Gini indices in all areas (both sampled and non-samples areas). 
We additionally generated 5000 samples of the area-wise parameters in the non-sampled areas and computed the posterior means. 
The spatial distribution of the estimates based on the proposed methods are presented in Figure~\ref{fig:Japan-est}. 
The figure shows that the proposed methods can produce spatially smoothed estimates of the average incomes and Gini indices not only in sampled areas but also non-sampled areas. 
This is the greatest advantage of the proposed method over the crude AML method. 
Comparing the results under the two different pair-wise priors, it is observed that PWL tends to produce spatially more smoothed estimates than PWD, and PWL detects some local hotspots.  
These phenomena may come from the properties of PWL; it forces income distributions in adjacent areas to be the same when a difference between the parameter values is small while it does not shrink the difference so much when the difference is significantly large.

\begin{table}[!htb]
\caption{
Posterior predictive loss for six models (combination of three parametric income distributions and two state space distributions).
\label{tab:PPL}
}
\begin{center}
\begin{tabular}{ccccccccccc}
\hline
 Distribution &  & \multicolumn{2}{c}{Log-normal}  &  \multicolumn{2}{c}{Singh-Madala} & \multicolumn{2}{c}{Dagum}  \\
 \# area-wise parameters && \multicolumn{2}{c}{2}  &  \multicolumn{2}{c}{3} & \multicolumn{2}{c}{3} \\
 \hline
State space distribution && PWD & PWL & PWD & PWL  & PWD & PWL \\
Posterior predictive loss && 1479 & 1479 & 679 & 676 & 705 & 2473 \\
 \hline
\end{tabular}
\end{center}
\end{table}

\begin{table}[!htb]
\caption{
Posterior mean and lower and upper values of $95\%$ credible intervals of structural parameters of the proposed models with the SM distributions and PWD and PWL latent structures.
\label{tab:para}
}
\begin{center}
\begin{tabular}{ccccccccccc}
\hline
 && \multicolumn{3}{c}{SM-PWD} && \multicolumn{3}{c}{SM-PWL} \\
 &  & Lower & Mean & Upper &  & Lower & Mean & Upper \\
 \hline
 $\mu_1$ &  & 0.656 & 0.665 & 0.674 &  & 0.645 & 0.660 & 0.675 \\
$\mu_2$ &  & 6.41 & 6.44 & 6.47 &  & 6.42 & 6.46 & 6.49 \\
$\mu_3$ &  & 0.836 & 0.856 & 0.880 &  & 0.847 & 0.876 & 0.907 \\
$\tau_1$ &  & 41.7 & 56.0 & 70.4 &  & 5.80 & 18.9 & 28.5 \\
$\tau_2$ &  & 2.53 & 3.76 & 5.33 &  & 1.04 & 4.84 & 10.2 \\
$\tau_3$ &  & 9.50 & 21.3 & 42.6 &  & 1.73 & 6.92 & 17.1 \\
$\la_1$ &  & 98.8 & 129 & 143 &  & 11.7 & 14.4 & 16.9 \\
$\la_2$ &  & 53.8 & 61.0 & 69.9 &  & ---  & ---  & ---  \\
$\la_3$ &  & 74.7 & 123 & 174 &  & ---  & ---  & --- \\
 \hline
\end{tabular}
\end{center}
\end{table}

\section{Conclusion and Discussion}\label{sec:conc}
We have developed an effective  method for estimating the area-wise spatial income distributions taking account of the geographical relationship.
Through the demonstration using the real and simulated datasets, the proposed method can produce spatially smoothed estimates of the income distributions and outperforms the crude area-wise separate method. 

While we are concerned only with the spatially varying income distributions in the present study, 
the grouped income data are available over different time periods, in which case, it would be desirable to estimate patio-temporal income distributions at the same time.
Our method can be generalized to a spatial-temporal method by introducing serial correlation of area-wise parameters, which would be an interesting future work.
Moreover, if we have additional information (e.g. sample mean) other than the grouped data, using the information would improve the estimation accuracy of parameters.
We may use the generalized method of moments (GMM) as considered in \cite{Haj2012}, especially the Bayesian GMM method \citep{Yin2006} to use in the framework of the proposed method. 
The detailed investigation is left to a future study.

\section*{Acknowledgements}
This work was supported by JSPS KAKENHI Grant Numbers 18K12754, 18K12757, 19K13667 and Japan Center for Economic Research.

\vspace{1cm}

\end{document}